\documentclass[twocolumn,pra]{revtex4}% Physical Review A

\usepackage{graphicx}% Include figure files
\usepackage{dcolumn}% Align table columns on decimal point
\usepackage{bm}% bold math

\begin{document}

\preprint{APS/123-QED}

\title{On the single mode approximation in spinor-1 atomic condensate}
\author{S. Yi$^1$, \"O. E. M\"ustecapl{\i}o\u{g}lu$^1$,
C. P. Sun$^{1,2}$, and L. You$^1$}

\affiliation{$^1$School of Physics, Georgia Institute of
Technology, Atlanta, GA 30332, USA}

\affiliation{$^2$Institute of Theoretical Physics, The Chinese
Academy of Sciences, Beijing 100080, People's Republic of China}
\date{\today}
%-------------------------------------------------------------------------------
\begin{abstract}
We investigate the validity conditions of the single mode
approximation (SMA) in spinor-1 atomic condensate when effects due
to residual magnetic fields are negligible. For atomic
interactions of the ferromagnetic type, the SMA is shown to be
exact, with a mode function different from what is
commonly used. However, the quantitative deviation is small
under current experimental conditions (for $^{87}$Rb atoms).
For anti-ferromagnetic
interactions, we find that the SMA becomes invalid in general.
The differences among the mean field mode functions for the
three spin components
are shown to depend strongly on the system
magnetization. Our results can be important for studies of beyond
mean field quantum correlations, such as fragmentation,
spin squeezing, and multi-partite entanglement.
\end{abstract}


\maketitle
%-------------------------------------------------------------------------------

Trapped atomic quantum gases have provided a remarkable testing ground
for quantum many-body theory \cite{Edwards}. Since the discovery of the
first atomic Bose-Einstein condensate \cite{bec}, mean field theory has
been applied with great success to these systems,
arguably because 1) low energy
atom-atom interaction can be simply parameterized by a s-wave
scattering length $a_{\rm sc}$ with atoms behave as
hard spheres of effective radii $a_{\rm sc}$; and 2) most current atomic gases are dilute with
densities $n$ satisfying $na_{\rm sc}^3\ll 1$ \cite{note1}.
Increasingly, theoretical and experimental attentions are
directed towards beyond mean field effects.
In this regard, spinor-1 atomic condensates have become
a proto-type system for many recent studies \cite{duan,pu,goldstein,ho2}.
Several interesting results have already been obtained,
e.g. multi-particle and continuous variable
type entanglement \cite{duan}, spin-mixing \cite{pu},
spinor four-wave mixing \cite{goldstein}, and super and coherent
fragmentation \cite{ho2}. The single mode
approximation (SMA) is often adopted for these studies
when a mean-field approach with a vectorial order parameter
becomes inappropriate \cite{ohmi,ho,ciobanu}.
Beyond mean field quantum effects have been
found both when there is no external fields
\cite{pu,goldstein} and when there is an external
magnetic or optical field \cite{pu3,pu4,pu5,raghavan,ueda}.
To justify the use of the SMA, earlier studies
often compared with solutions of the coupled
Gross-Pitaevskii (GP) equation for the different spin components
and enforced an upper limit on the number of atoms \cite{duan,pu3}.
While there is not a generally adopted limit,
it is typically estimated that $N$
should be less than $10^4$, a rather small number
for current experiments.

In this paper, we investigate the validity conditions of the SMA
in spinor-1 atom condensate \cite{kurn,mike}.
Our initial aim was to provide a reliable
thermodynamic phase diagram for a trapped spinor-1
atomic gas \cite{machida}. Surprisingly,
interesting zero temperature results from the
coupled GP equations reveal
intricate relationships of the mode functions for the three spin components
due to the constraint on the system magnetization.

We consider a spinor-1 atomic condensate
in the absence of an external magnetic field.
As partitioned by Law {\it et al.} \cite{pu},
the system Hamiltonian, $H$, separates into a
symmetric part (under spin exchange)
\begin{eqnarray}
H_S=\int d\vec r\, \left(\Psi_\alpha^\dag{\cal
L}_{\alpha\beta}\Psi_\beta+{c_0\over
2}\Psi_\alpha^\dag\Psi_\beta^\dag\Psi_\beta\Psi_\alpha\right),
\label{hs}
\end{eqnarray}
with ${\cal L}_{\alpha\beta}
=-{\hbar^2}\nabla^2/2M+V_{\rm ext}$,
and an asymmetric part
\begin{eqnarray}
H_A&=&{c_2\over 2}\int d\vec r\,
\Psi_\alpha^\dag\left(F_\eta\right)_{\alpha\beta}\Psi_\beta
\Psi_\mu^\dag\left(F_\eta\right)_{\mu\nu}\Psi_\nu,\label{ha}
\end{eqnarray}
where $\Psi_\alpha$ ($\alpha=0,\pm$) denotes the annihilation
field operator for the $\alpha$-th component.
$F_{\eta=x,y,z}$ are the spin 1 matrix representation, and a
summation over repeated indices is assumed in Eqs. (\ref{hs}) and
(\ref{ha}). The external trapping potential $V_{\rm ext}(\vec r)$
is spin-independent as in an far off-resonant optical
dipole force trap (FORT) which makes atomic spinor degrees of freedom
completely accessible. The pair
interaction coefficients are $c_0=4\pi\hbar^2(a_0+2a_2)/3M$ and $
c_2=4\pi\hbar^2(a_2-a_0)/3M$, with $a_0$ ($a_2$) the s-wave
scattering length for two spin-1 atoms in the combined symmetric
channel of total spin $0$ ($2$). The only state changing
collision in Eq. (\ref{ha}) occurs through the
coupling $\Psi_0^\dag\Psi_0^\dag\Psi_+\Psi_-+h.c.$, which
conserves the system magnetization ${\cal M}=\int
d\vec r\langle F_z\rangle=\int d\vec r
[\Psi_+^\dag\Psi_+-\Psi_-^\dag\Psi_-]$. ${\cal M}$-changing
inelastic (``bad") collisions occur at a much longer time scale
as compared with a condensate's typical lifetime, therefore
are excluded here as in all previous studies. Although the real
time dynamics governed
by $H_S+H_A$ conserves the total atom number
$N=\int d\vec r
[\Psi_+^\dag\Psi_++\Psi_0^\dag\Psi_0 +\Psi_-^\dag\Psi_-]$ and ${\cal M}$,
the ground state obtained from a
global minimization of $H_S+H_A$ is not automatically
guaranteed to have the same $N$ and ${\cal M}$.
We therefore introduce separate Lagrange multipliers ${\cal B}$
to guarantee the conservation of ${\cal M}$ and
the chemical potential $\mu$ to conserve $N$.
The ground state is then determined by a minimization
of the free energy functional ${\cal F}=H_S+H_A-\mu N-{\cal B}{\cal M}$.
Mathematically, this task turns out to be highly nontrivial.
In fact, most previous discussions on spinor-1 condensates
did not minimize $H$ under the constraint of a conserved $\cal M$.
Therefore, their resulting ground states are the global ground
states that can only be reached if the system can coherently
adjust its initial ${\cal M}$ value.
Such a situation is inconsistent with current experiments.

One of the strongest physics support for the SMA comes
from the fact that $a_0\sim a_2$ for a spinor-1 ($^{87}$Rb) condensate.
This gives rise to $|c_2|\ll |c_0|$ \cite{pu,ho}. Thus
$H_A$ is much smaller as compared with $H_S$, and
can be considered as a perturbation by assuming the SMA
\begin{eqnarray}
\Psi_\alpha(\vec r)=a_\alpha\phi_{\rm SMA}(\vec r), \qquad
\alpha=0,\pm, \label{sma}
\end{eqnarray}
i.e. with a common mode function $\phi_{\rm SMA}(\vec r)$
(normalized to $1$). The Fock state
boson operators $a_{\alpha}$ satisfy
$[a_{\alpha},a_{\gamma}^{\dag}]=\delta_{\alpha\gamma}$,
$[a_{\alpha},a_{\gamma}]=0$. $\phi_{\rm SMA}(\vec r)$ is determined
from $H_S$ alone (without $H_A$) according to \cite{pu}
\begin{eqnarray}
\left[-{\hbar^2\nabla^2\over 2M}+V_{\rm ext}+c_0N|\phi_{\rm
SMA}|^2\right]\phi_{\rm SMA}(\vec r)=\mu\phi_{\rm SMA}(\vec r).
\label{esma}
\end{eqnarray}
It shares similar physics of the often used spin-charge
separation in condensate matter systems.
Since its introduction, the SMA has been used frequently
\cite{duan,pu,goldstein,ho2,ueda2}. Notable exception
is the work by Ueda \cite{ueda3}, who went beyond the SMA
by studying a translational invariant system with the use
of a plane wave basis. Correlations between spatial
and spinor degrees of freedom were then shown to lead to
effects associated with density waves and spin waves.
For a trapped system as studied here, the use of a plane
wave basis becomes inappropriate.

The same SMA is sometimes also used in a spin $1/2$ system by
assuming $\phi_0(\vec r)=\phi_1(\vec r)$ \cite{poulsen,sorensen,zoller,you}.
This is less critical as the resulting Hamiltonian $\propto J_z^2$
remains of the same symmetry group in the Schwinger boson
representation, although with a different coefficient and
the presence of additional linear terms in $J_{\mu}$.
The validity of the SMA in this case
has been tested recently using the
rigorous positive P-approach \cite{poulsen,sorensen}.
\begin{figure}
\includegraphics[width=2.7in]{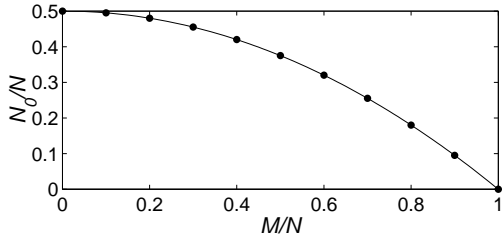}
\caption{The ${\cal M}$ dependence of $N_0$ in the ferromagnetic
case. The solid line shows $N_0/N=(1-{\cal M}^2/N^2)/2$, while the
dots are numerical results. The agreement is remarkable.\\[-24pt]}
\label{ferrn0}
\end{figure}

For a spinor-1 condensate, however, complications arise
when spin component mode functions are taken
to be different. The effective Hamiltonian thus obtained
contains no angular momentum symmetry at all in its
corresponding Schwinger boson representation. This
naturally calls for a critical investigation of the SMA.
To check the validity of SMA, we start with the
mean field and find separate spin component mode functions
$\langle\Psi_\alpha\rangle=\Phi_\alpha$ (at zero
temperature). The dynamics of $\Phi_\alpha$ for the
ground state is governed by $H_S+H_A$, which obeys the
following coupled GP equation
\begin{eqnarray}
i\hbar\dot{\Phi}_+&=&\left[{\cal H}-{\cal B}
+c_2(n_++n_0-n_-)\right]\Phi_++c_2\Phi_0^2\Phi_-^*,\nonumber\\
i\hbar\dot{\Phi}_0&=&\left[{\cal H}
+c_2(n_++n_-)\right]\Phi_0+2c_2\Phi_0^*\Phi_+\Phi_-, \label{scht0}\\
i\hbar\dot{\Phi}_-&=&\left[{\cal H}+{\cal B}+
c_2(n_-+n_0-n_+)\right]\Phi_-+c_2\Phi_0^2\Phi_+^*,\nonumber
\end{eqnarray}
with ${\cal H}={-\hbar^2\nabla^2/2M }+V_{\rm ext}+c_0n$,
$n_{\alpha}=|\Phi_\alpha|^2$, and $n=n_++n_0+n_-$. We have
developed a reliable numerical algorithm based on propagating Eq.
(\ref{scht0}) in imaginary time ($it$) that converges to the
ground state while maintaining the conservation of both
$N$ and $\cal M$. We take the initial
wave function to be a complex Gaussian with a constant velocity,
i.e., $e^{-(x^2/2q_x^2+y^2/2q_y^2+z^2/2q_z^2)}e^{-i\vec k\cdot\vec
r},$ where $q_x$, $q_y$, $q_z$, and $\vec k$ are adjustable
parameters that shall not affect the final converged solution.
In the simplest case for the ground state, we assume
$\Phi_\alpha(\vec r)=|\Phi_\alpha(\vec r)|e^{i\theta_\alpha}$ with
$\theta_\alpha$ a global phase independent of $\vec r$.
Then only the relative phase $\Delta=2\theta_0-\theta_+-\theta_-$
shows up in ${\cal F}$ with a term $\propto
c_2|\Phi_+\Phi_-\Phi_0^2|\cos\Delta$. This gives
$\Delta=$ $0$ (for $c_2<0$) or $\pi$ (for $c_2>0$)
when ${\cal F}$ is minimized \cite{new}, a conclusion also verified
by numerical calculations. As first
stated by Ho \cite{ho}, the spinor-1 condensate Hamiltonian
$H=H_S+H_A$ is invariant under gauge transformation $e^{i\theta}$
and spin rotations ${\cal
U}(\alpha,\beta,\tau)=e^{-iF_z\alpha}e^{-iF_y\beta}e^{-iF_z\tau}$.
For the ground state that conserves $\cal M$, however, the spin
rotation symmetry is reduced to the subgroup SO(2) generated by
$e^{-iF_z\alpha}$. Thus irrespective of the signs of $c_2$, a
transformation of the form
$e^{-i\theta_0}e^{-iF_z(\theta_+-\theta_-)/2}$ can always reduce
a complex solutions to a real one \cite{new}.
\begin{figure}
\includegraphics[width=3.2in]{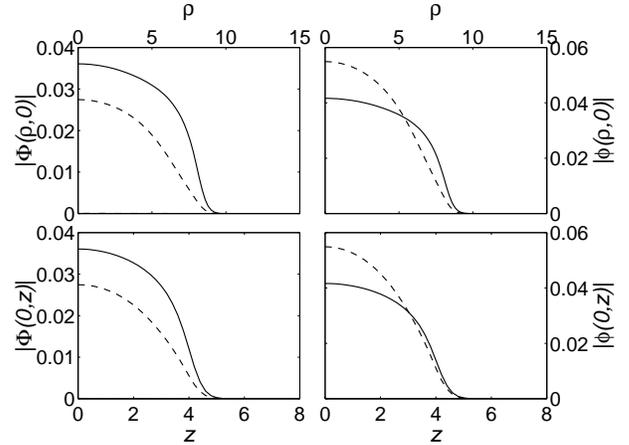}
\caption{The original (left column) and the renormalized (right
column) wave functions along radial (upper panel) and axial (lower
panel) directions for the $+$ (solid line) and the $-$ (dashed
line) spin components. Other parameters are $N=3.16\times 10^5$,
${\cal M}/N=0.5$, and $\lambda=2$. All lengths are in units of
$\sqrt{\hbar/m\omega_r}$.\\[-24pt]} \label{wl2n1e6m05}
\end{figure}

When ${\cal B}=0$ as for
ferromagnetic interactions with any values of magnetization
${\cal M}\leq N$ or for anti-ferromagnetic interactions with
${\cal M}=0$, we find $|\phi_+|\equiv|\phi_-|$ from the
symmetry of Eq. (\ref{scht0}). We then rescale the wave function
$\phi_\alpha=\Phi_\alpha/\sqrt{N_\alpha}$ such that $\phi_\alpha$ is
normalized to 1 ($\int d\vec r |\Phi_\mu(\vec r)|^2=N_\mu$, the
number of atoms in $\mu$-th component), the asymmetric
interaction energy then becomes
\begin{eqnarray}
E_A &=&{c_2\over 2}\int d\vec r
\left[\left(N_+|\phi_+|^2-N_-|\phi_-|^2\right)^2\right.\nonumber\\
&+&2N_+N_0|\phi_+|^2|\phi_0|^2
+2N_-N_0|\phi_-|^2|\phi_0|^2\nonumber\\
&+&\left.4N_0(N_+N_-)^{1/2}|\phi_0|^2|\phi_+||\phi_-|\cos\Delta\right].\label{ea}
\end{eqnarray}
For ferromagnetic interactions ($c_2<0$ and $\Delta=0$), we
thus prove in general that $E_A$ is minimized when
\begin{eqnarray}
|\phi_+|=|\phi_0|=|\phi_-|=|\phi|,\label{m0}
\end{eqnarray}
and $N_0/N=(1-{\cal M}^2/N^2)/2$. The latter result (independent of all
other parameters) was first obtained in Ref. \cite{pu3} assuming the
SMA, i.e. essentially assuming Eq. (\ref{m0}). Our numerical
solutions closely follow this as shown in Fig. \ref{ferrn0}.
For anti-ferromagnetic interactions ($c_2>0$),
${\cal B}=0$ holds only when ${\cal M}=0$.
In this case, using $\Delta=\pi$, we prove in general that
$E_A$ is minimized to zero under Eq. (\ref{m0}), while $N_0$ can
be any value $\le N$ \cite{pu3}.

For anti-ferromagnetic interactions (${\cal M}\neq 0$),
we find that mode functions for the three spin
components are different (see Fig. \ref{wl2n1e6m05}).
Further analysis show that $E_A$ is minimized if $N_0=0$ \cite{pu3}.
\begin{figure}
\centering
\includegraphics[width=3.25in]{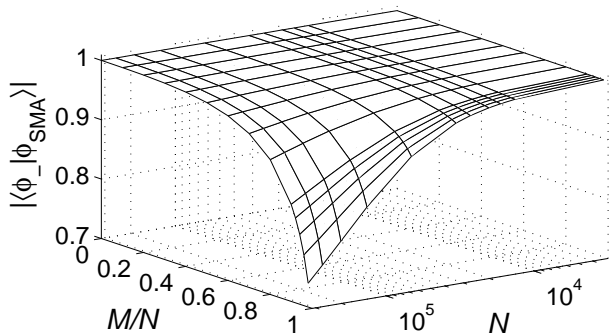}
\caption{${\cal M}$ and $N$ dependence of the overlap
integral $|\langle\phi_-|\phi_{\rm SMA}\rangle|$ for $^{23}$Na
atoms ($\lambda=1$). $|\langle\phi_+|\phi_{\rm SMA}\rangle|$
is essentially 1 to within $\pm0.001$ in the same region.\\[-24pt]} \label{oversma}
\end{figure}

We now discuss the relationship of Eq. (\ref{m0}) to
the SMA Eq. (\ref{esma}). We note that the validity of Eq. (\ref{m0})
(including $H_A$) is in fact not equivalent to the validity of the SMA
(excluding $H_A$). For ferromagnetic interactions, with Eq.
(\ref{m0}) and the relation between $N_0$ and ${\cal M}$,
equation (\ref{scht0}) simplifies to
\begin{eqnarray}
\left[-{\hbar^2\nabla^2\over 2M}+V_{\rm
ext}+(c_0+c_2)N|\phi|^2\right]\phi(\vec r)=\mu\phi(\vec
r).\label{fgp}
\end{eqnarray}
This shows that $\phi(\vec r)$ is independent of ${\cal M}$,
and its deviation from $\phi_{\rm SMA}$ comes only from the $c_2$
term. This result can in fact be easily understood. Since
$c_0+c_2=4\pi\hbar^2 a_2/M$, $\phi(\vec r)$ of Eq. (\ref{fgp}) is
simply the ground state of the GP equation for an atomic
scattering length of $a_2$. In a ferromagnetic state, atomic spins
are aligned locally. Two such atoms ($F_{1,2}=1$) only collide
in the symmetric total spin $F=2$ channel. For
quantitative results, we compared $|\langle\phi|\phi_{\rm
SMA}\rangle|$ for $^{87}$Rb atoms with $a_0=101.8\,a_B$ and
$a_2=100.4\,a_B$ \cite{heinzen}, ($a_B$ the Bohr radius).
Other assumptions are: typical radial trap frequency $\omega_r=2\pi\times
10^3\,$(Hz), axial trap frequency $\omega_z=\lambda\omega_r$, and
$\lambda=0.1$, $1$, and $10$. We also took $N=3.16\times 10^5$ and
varied the total magnetization ${\cal M}/N$ from $0$ to $1$. Under
these conditions, we find uniformly that $|\langle\phi|\phi_{\rm
SMA}\rangle|\approx 1$ essentially because $|c_2|\ll |c_0|$.

\begin{figure}
\includegraphics[width=2.7in]{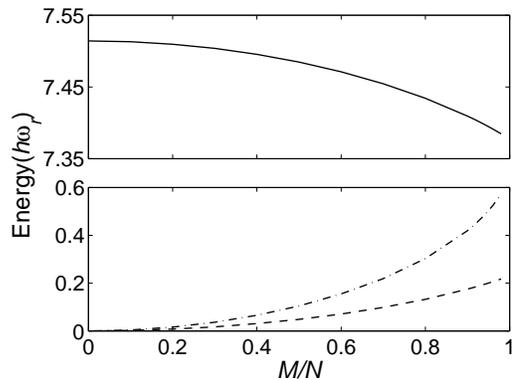}
\caption{The ${\cal M}$ dependence of energy components (in units of
$\hbar\omega_r$) $E_{c_0}$ (solid line), $E_A$ (dashed line), and
${\cal BM}$ (dash-dotted line) for $^{23}$Na atom at $\lambda=1$
and $N=3.16\times 10^5$.\\[-24pt]} \label{energy}
\end{figure}

\begin{figure}[h]
\includegraphics[width=2.7in]{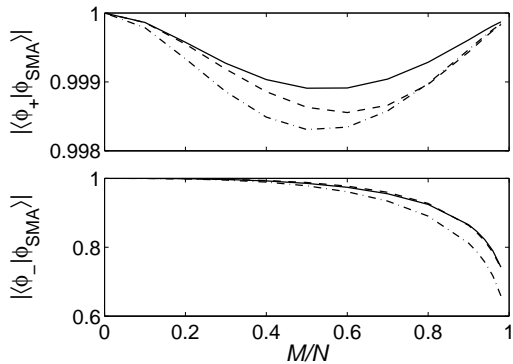}
\caption{The overlap $|\langle\phi_\pm|\phi_{\rm SMA}\rangle|$
for $\lambda=0.1$ (solid line), $1$ (dashed line), and $10$
(dash-dotted line). } \label{osmalmd}
\end{figure}

For anti-ferromagnetic interactions ($c_2>0$), two special cases
arise: 1) when ${\cal M}=0$, using Eq. (\ref{scht0}), we prove
that $|\phi_\alpha|=|\phi_{\rm SMA}|$, which means the SMA is exact
in this case; 2) when ${\cal M}=N$, $\phi_+$
satisfies the same equation as Eq. (\ref{fgp}), so its derivation
from the SMA only originates from the $c_2$ term. For $^{23}$Na atoms,
we use $a_0=50\,a_B$ and $a_2=55\,a_B$ \cite{tiemann} as an example in
this case. Other parameters used are the same as in the
ferromagnetic case. Since $N_0=0$, we consider only the $\pm$
components. Figure \ref{wl2n1e6m05} shows the original and
renormalized wave function for $N=3.16\times 10^5$, ${\cal
M}/N=0.5$, and $\lambda=2$. We see clearly that $\phi_+$ and $\phi_-$
are different. Figure \ref{oversma} shows the magnetization and
atom number dependence of $|\langle\phi_-|\phi_{\rm SMA}\rangle|$
for a spherical trap. Since the $+$ component contains the
majority number of atoms, it is natural to find
$|\langle\phi_+|\phi_{\rm SMA}\rangle|\approx 1$. The value of
$|\langle\phi_+|\phi_{\rm SMA}\rangle|$ at ${\cal M}\approx N$
also indicates that the deviation contributed by $c_2$ alone is
also small for $^{23}$Na atoms. While for
$|\langle\phi_-|\phi_{\rm SMA}\rangle|$, we see it
becomes as low as
$0.75$ when $N=3.16\times 10^5$ and when ${\cal M}$ approaches $N$.
To distinguish the different sources of deviations, we plot
$E_{c_0}=(c_0/2)\int d\vec r n^2$, $E_A$,
and ${\cal BM}$ in Fig. \ref{energy}. We see that the ${\cal BM}$ term
contributes the most. In Fig. \ref{osmalmd}, the overlap integral
is shown to also depend on the trap aspect ratio $\lambda$.

Finally, we discuss the implications of our result on the
macroscopic alignment of the total spin of a spinor condensate.
For ferromagnetic interactions, the spatial distribution of the
total spin $\langle\vec F(\vec r)\rangle\equiv\sum_{\alpha\beta}
\Phi_\alpha^*(\vec r) \vec F_{\alpha\beta}\Phi_{\beta}(\vec r)$ is
found to be pointed along the same direction, i.e. independent of
the spatial coordinates. Using $\Delta=0$, $N_0=N(1-{\cal
M}^2/N^2)/2$, and $N_\pm=N(1\pm{\cal M}/N)^2/4$, it can be
expressed as
\begin{eqnarray}
\langle\vec F(\vec r)\rangle =|\phi(\vec
r)|^2\left(\begin{array}{c}
\sqrt{N^2-{\cal M}^2}\,\cos (\theta_+-\theta_0)\\
-\sqrt{N^2-{\cal M}^2}\,\sin (\theta_+-\theta_0)\\
{\cal
M}\end{array}\right).
\end{eqnarray}

For anti-ferromagnetic interactions, we find
\begin{eqnarray}
\langle\vec F(\vec r)\rangle
=\left(\begin{array}{c}0\\
0\\
N_+|\phi_+(\vec r)|^2-N_-|\phi_-(\vec r)|^2\end{array}\right),
\end{eqnarray}
a state with all spins aligned in the $\pm z$ direction. It
reduces to $\langle\vec F(\vec r)\rangle=\vec 0$ for ${\cal M}=0$.

To conclude, we presented both analytic and numerical studies of
the validity of the SMA. We find that deviations of the
ground state solution from the $\phi_{\rm SMA}(\vec r)$
come from two sources: the $c_2$ or the ${\cal B}$ (due to
conservation of ${\cal M}$) term. For ferromagnetic
interactions, the only source is the $c_2$ term, which is negligible
for $^{87}$Rb atoms. One can therefore
safely use the SMA. For anti-ferromagnetic interactions,
if ${\cal M}=0$, $\phi_{\rm SMA}$
becomes the exact ground state wave function;
For ${\cal M}>0$, however, one
can still use the $\phi_{\rm SMA}$ for $\phi_+$, but
$\phi_-$ differs significantly
if both $N$ and ${\cal M}$ are large.
In this case the ${\cal BM}$ term contributes the most to the deviation.
Our conclusions from this study apply to the ground states of a spinor condensate.
For dynamic problems Ref. \cite{pu3}, the SMA may become worse.
Our study suggests that instead of making the SMA as in Eq. (\ref{sma}),
an improved SMA could consist of $\Psi_\mu=a_\mu\phi_\mu(\vec r)$,
where the mean-field solution
$\Phi_\mu(\vec r)$ and its associated effective spin mode
function $\phi_\mu={\Phi_\mu(\vec r)/\sqrt{N_\mu}}$
are obtained under the constraints of
conserved $N$ and ${\cal M}$. Such an approach can be important
in studying beyond mean field quantum correlations. In a
forthcoming article, we will report some results
on condensate fragmentation.

In summary, we have presented a detailed investigation of the SMA
for a spinor-1 condensate and pointed out interesting structures
of its ground state for both ferromagnetic
and anti-ferromagnetic interactions.\\

This work is supported by the NSF grant No. PHY-9722410
and by a grant from the National Security Agency (NSA),
Advanced Research and Development Activity (ARDA), and the Defense
Advanced Research Projects Agency (DARPA) under Army Research Office
(ARO) Contract No. DAAD19-01-1-0667. Partial support
from the NSF of China is also acknowledged.

\end{document}